\documentclass{article}

\usepackage[preprint]{neurips_2025}

\usepackage[utf8]{inputenc} % allow utf-8 input
\usepackage[T1]{fontenc}    % use 8-bit T1 fonts
\usepackage{booktabs}       % professional-quality tables
\usepackage{amsfonts}       % blackboard math symbols
\usepackage{nicefrac}       % compact symbols for 1/2, etc.
\usepackage{microtype}      % microtypography
\usepackage{xcolor}         % colors
\usepackage{graphicx}
\usepackage{subcaption}  % if you want subfigures
\usepackage{float}       % improved figure placement
\usepackage{tabularx}
\usepackage{amsmath}
\usepackage{caption}
\usepackage{multirow}    % multi-row cells
\usepackage{array}       % column formatting
\usepackage{adjustbox}   % scaling tables if needed
\usepackage{listings}
\definecolor{myblue}{rgb}{0,0.3,0.6}
\usepackage[colorlinks=true,linkcolor=myblue,urlcolor=myblue,citecolor=myblue,anchorcolor=myblue]{hyperref}     

\lstdefinelanguage{json}{
  basicstyle=\ttfamily\small,
  numbers=left,
  numberstyle=\tiny,
  stepnumber=1,
  numbersep=8pt,
  showstringspaces=false,
  breaklines=true,
  stringstyle=\color{blue},
  literate=
   *{0}{{{\color{black}0}}}{1}
    {1}{{{\color{black}1}}}{1}
    {2}{{{\color{black}2}}}{1}
    {3}{{{\color{black}3}}}{1}
    {4}{{{\color{black}4}}}{1}
    {5}{{{\color{black}5}}}{1}
    {6}{{{\color{black}6}}}{1}
    {7}{{{\color{black}7}}}{1}
    {8}{{{\color{black}8}}}{1}
    {9}{{{\color{black}9}}}{1}
}

\title{The \textit{Name-Free Gap}: Policy-Aware Stylistic Control in Music Generation}

\author{%
  Ashwin Nagarajan \\
  University of California, Santa Cruz\\
  \texttt{asnagara@ucsc.edu} \\
  \And
  Hao-Wen Dong \\
  University of Michigan\\
  \texttt{hwdong@umich.edu} \\
}

\begin{document}

\maketitle

\begin{abstract}

Text-to-music models capture broad attributes such as instrumentation or mood, but fine-grained stylistic control remains an open challenge. Existing stylization methods typically require retraining or specialized conditioning, which complicates reproducibility and limits policy compliance when artist names are restricted. We study whether lightweight, human-readable modifiers sampled from a large language model can provide a policy-robust alternative for stylistic control. Using MusicGen-small, we evaluate two artists: Billie Eilish (vocal pop) and Ludovico Einaudi (instrumental piano). For each artist, we use fifteen reference excerpts and evaluate matched seeds under three conditions: baseline prompts, artist-name prompts, and five descriptor sets. All prompts are generated using a large language model. Evaluation uses both VGGish and CLAP embeddings with distributional and per-clip similarity measures, including a new min-distance attribution metric. Results show that artist names are the strongest control signal across both artists, while name-free descriptors recover much of this effect. This highlights that existing safeguards such as the restriction of artist names in music generation prompts may not fully prevent style imitation. Cross-artist transfers reduce alignment, showing that descriptors encode targeted stylistic cues. We also present a descriptor table across ten contemporary artists to illustrate the breadth of the tokens. Together these findings define the \textit{name-free gap}, the controllability difference between artist-name prompts and policy-compliant descriptors, shown through a reproducible evaluation protocol for prompt-level controllability.

\end{abstract}

\section{Introduction}

Recent progress in text-to-music generation has enabled models such as MusicLM \citep{agostinelli2023musiclm}, MusicGen \citep{copet2023musicgen}, and MusicLDM \citep{chen2023musicldm} to produce coherent audio from natural language prompts. While these systems capture broad attributes like instrumentation or mood, they offer limited control over the stylistic cues that define an artist’s identity. Stylization methods typically require costly retraining or additional adapters, which hinders reproducibility for users. In addition, stylization using artist names often conflicts with deployment policies of text-to-music platforms, motivating the need for policy-compliant alternatives.

\begin{figure}[t]
    \centering
    \includegraphics[width=0.9\linewidth]{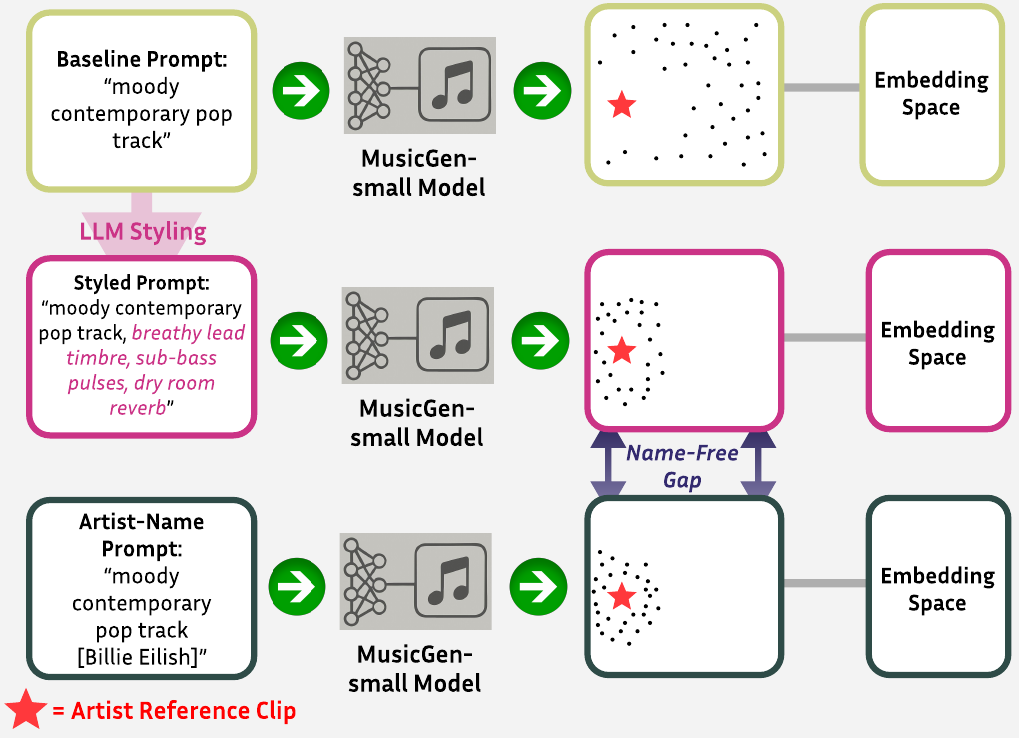}
    \caption{\textbf{Prompt-level controllability and the \textit{name-free gap}:} A baseline prompt (top) produces dispersed generations in embedding space relative to the artist reference clip. Adding LLM-sampled descriptors shifts samples closer to the references, while artist-name prompts achieve the strongest alignment. The difference between descriptor and artist-name conditions defines the \textit{name-free gap}.}
    \label{fig:teaser}
\end{figure}

We ask whether lightweight, human-readable modifiers sampled from a large language model can provide a policy-compliant alternative for stylistic control. Figure~\ref{fig:teaser} outlines this setup. These descriptors are interpretable, require no model modifications, and can be generated for any artist without using their audio, which avoids copyright issues and supports reproducibility. To test this idea, we evaluate two artists: Billie Eilish, a vocal-driven pop artist, and Ludovico Einaudi, a piano-based instrumental composer. Our contributions include: (1) a reproducible protocol for evaluating prompt-level controllability, (2) baselines and controls including artist-name prompts, min-distance attribution, and cross-artist transfer, and (3) evidence that LLM-sampled descriptors induce measurable, artist-specific shifts while recovering much of the effect of artist names. Together these results define the \textit{name-free gap}, highlight that our descriptors can achieve similar stylistic effects to artist names, and motivate strategies for controllable music generation that remain accessible under platform policies. For reproducibility, the appendix provides links to GitHub, HuggingFace, and a website with all code, manifests, embeddings, generated audio samples, and demos.\footnote{\url{https://artisticstyles.github.io/music-style-control-demo/}}

\section{Related Work}

\paragraph{Text-to-music generation.}  
Recent models such as MusicLM \citep{agostinelli2023musiclm} generate coherent audio from text, and MusicGen \citep{copet2023musicgen} shows that a single autoregressive model can support both text and melody conditioning. However, controllability remains limited when relying only on free-form prompts.

\paragraph{Stylization.}  
Prior work on Stylization often relies on finetuned embeddings or learned tokens, including textual inversion for images \citep{gal2022textualinversion}, prompt-to-prompt edits \citep{hertz2022prompttoprompt}, and style tokens for speech \citep{wang2018gst}. In music, Rouard et al. \citep{rouard2024audioconditioning} introduce discrete bottleneck features that require model adaptation. Our approach instead uses lightweight, human-readable descriptors sampled from a language model, which do not require optimization and provide a policy-compliant alternative to artist names.

\paragraph{Evaluation.}  
Generative audio is typically evaluated in embedding space. VGGish underpins Fréchet Audio Distance (FAD) \citep{hershey2017vggish,kilgour2019fad}, while CLAP provides broader joint audio text representations \citep{elizalde2022clap,microsoft2023clap}. We evaluate using FAD and a new min-distance attribution metric in both spaces, along with cross-artist validation as a specificity control, while noting that embedding-based scores remain imperfect proxies for perception \citep{huang2025aligningttm}.

\section{Methods}

\subsection{Data and Generation}
We test whether lightweight textual modifiers can steer generated outputs toward the distribution of a target artist.  
For each artist, we use a large language model to generate a baseline prompt that captures broad instrumentation and mood. For example, “moody contemporary pop track” for Billie Eilish and “instrumental track with gentle dynamics” for Ludovico Einaudi.    
Candidate descriptors are proposed by a large language model, with five independent sets sampled per artist.  
Each styled prompt is formed by appending one of these sets to the baseline. An additional baseline appends only the artist name, providing a control for trivial name-based stylization.  
Table~\ref{tab:tokens} shows sample descriptors for ten artists, showing that the language model produces concise tokens reflecting stylistic traits across diverse genres. For the prompt used to query the LLM and other specific method details, refer to Appendix~\ref{appendix:sec:detailed-methods}.

Artists are represented by fifteen 15-second reference excerpts, resampled to 32 kHz and used only for embedding-based evaluation.  
All experiments use the publicly released \texttt{musicgen-small} checkpoint without modification.  
There are seventy total generated audio files per artist (fifty styled clips, ten baseline, and ten artist-name baseline clips).

\begin{table}[t]
\centering
\setlength{\tabcolsep}{4pt}      % tighter horizontal padding
\renewcommand{\arraystretch}{1.05} % slightly tighter rows
\footnotesize
\begin{tabularx}{\linewidth}{@{}p{2.8cm} X@{}}
\toprule
\textbf{Artist} & \textbf{Representative LLM-sampled descriptors} \\
\midrule
\textit{Billie Eilish}        & breathy lead timbre; sub-bass pulses; dry room reverb \\
\textit{Ludovico Einaudi}     & delicate piano timbre; string ensemble pad; intimate mix \\
\midrule
Zayn Malik           & airy vocal texture; deep sub bass; minimal trap rhythm \\
The Weeknd           & airy falsetto timbre; analog synth bass; dark reverberant space \\
Tyler, The Creator   & gritty vocal texture; distorted synth bass; experimental offbeat groove \\
Bruno Mars           & smooth falsetto tone; retro synth stabs; groovy disco bass \\
Taylor Swift         & warm lead timbre; acoustic guitar strums; intimate close space \\
Ariana Grande        & silky lead timbre; glossy synth chords; wide reverb space \\
The Beatles          & warm lead timbre; jangly electric guitar; tight backbeat groove \\
Drake                & hazy vocal timbre; 808 bass hits; moody trap groove \\
\bottomrule
\end{tabularx}
\caption{\textbf{Sample LLM descriptors:} The table shows the two evaluated artists and eight additional contemporary artists, each with three LLM-generated descriptors.  In the experiments, all three descriptors are appended to the baseline prompt to form styled prompts; see Tables~\ref{tab:prompts-billie} and \ref{tab:prompts-einaudi} for the exact prompt strings used.}
\label{tab:tokens}
\vspace{-2mm}
\end{table}

\subsection{Evaluation and Controls}
Evaluation is performed in two embedding spaces to control for embedding choice: VGGish \citep{hershey2017vggish} and CLAP \citep{elizalde2022clap,microsoft2023clap}.  
We focus on three measures: (i) FAD to quantify distributional similarity, (ii) min-distance attribution as a per-clip similarity metric, and (iii) cross-artist validation as a specificity control.

\paragraph{Min-distance attribution.}  
For each generated clip embedding $\mathbf{g}_i$ and reference set $\{\mathbf{r}_1,\dots,\mathbf{r}_N\}$, we compute  
\begin{equation}
d_{\text{min}}(\mathbf{g}_i) = \min_{j=1}^N \left( 1 - \frac{\mathbf{g}_i \cdot \mathbf{r}_j}{\|\mathbf{g}_i\| \|\mathbf{r}_j\|} \right),
\end{equation}
the cosine distance to the nearest reference.  
For each condition $c$, we report the raw median of these min-distance values:
$\operatorname{Median}(d_c).$
Lower values indicate closer alignment of generated clips to the target reference style.

\paragraph{Specificity controls.}  
To test whether descriptors are artist-specific rather than generically helpful, we adopt a cross-artist validation setup.  
Styled prompts sampled for one artist are evaluated against reference sets for both artists, and per-clip cosine similarities to the reference centroid are averaged. Improvements are computed relative to the baseline as  
\begin{equation}
\Delta = \operatorname{mean}(\text{styled centroid similarities}) - \operatorname{mean}(\text{baseline centroid similarities}).
\end{equation}
Higher positive diagonal values (same-artist) and negative or lower off-diagonal values (cross-artist) indicate that descriptors encode targeted stylistic cues rather than generic quality gains.

\section{Results and Discussion}

\begin{figure}[t]
\centering
\includegraphics[width=0.995\linewidth]{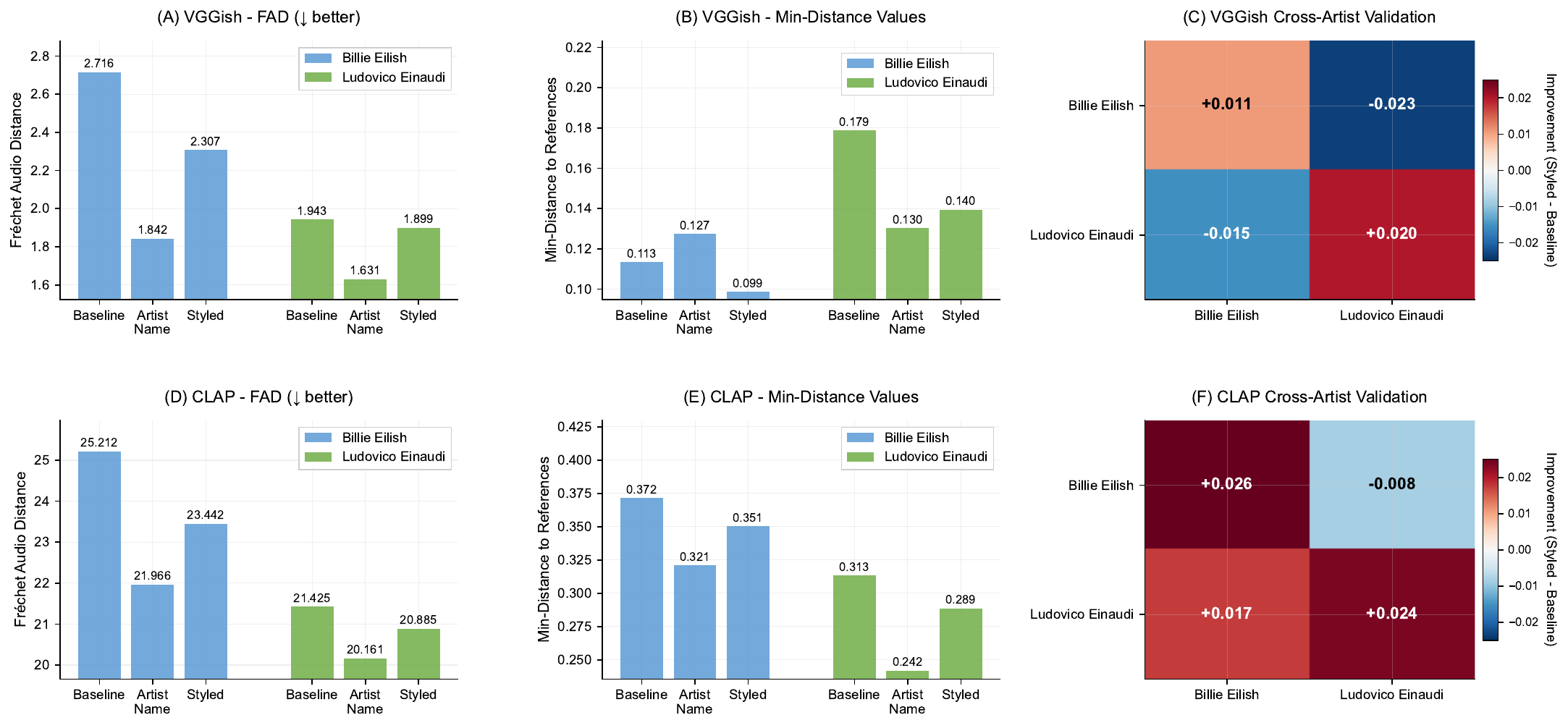}
\caption{\textbf{Dual embedding analysis:} 
(A,D): Comparison of both CLAP and VGGish FAD for both artists. (B, E): Comparison of min-distance values in CLAP and VGGish for both artists. (C, F): Cross-artist validation in both CLAP and VGGish.}
\label{fig:mainresults}
\end{figure}

\paragraph{Main Findings.}
Across both artists and embedding spaces, artist-name prompts provide the strongest improvements. Descriptor-augmented prompts consistently improve over the baseline and close a substantial portion of the \textit{name-free gap}. Panels A and D in Figure~\ref{fig:mainresults} show that artist names achieve the lowest FAD, with descriptors narrowing the difference. Panels B and E show reduced median min-distance under both artist-name and descriptor conditions relative to the baseline, though for Billie Eilish in Panel B the artist-name prompts actually perform slightly worse than the baseline, while the descriptor condition achieves the best result.

\paragraph{Specificity and Controls.}
Panels~C and F test specificity: diagonal cells (same artist) are higher positive values, while off-diagonal cells (cross artist) are negative or lower positive values, indicating that descriptors capture targeted stylistic cues rather than generic quality gains.

\paragraph{Discussion.}
Our findings show that prompt-level modifiers provide a lightweight means of steering generative music without retraining. Artist names act as strong style carriers, and name-free descriptors recover a substantial portion of this effect, which suggests that restricting artist names alone may not fully prevent style imitation.  Limitations include evaluation on two artists, short 15 second reference clips, and reliance on embedding-based proxies that do not replace listening tests; results are also sensitive to embedding choice, with VGGish and CLAP not always agreeing.

\section{Contributions and Future Works}

This work contributes a reproducible framework for evaluating prompt-level controllability in generative audio, combining matched seeds, fixed references, and evaluation in two embedding spaces, with controls such as artist-name prompts and cross-artist transfer, and metrics including FAD and min-distance attribution. Using this framework, we quantify the \textit{name-free gap} between artist-name prompts and policy-compliant descriptors, and outline how it can organize future research questions. While demonstrated with two artists, the approach is model-agnostic and applies broadly across music and related generative audio tasks. Future directions include expanding the artist and genre set, testing longer clips and musical structure, incorporating human listening studies, exploring alternative embeddings and robustness analyses, and developing systematic descriptor design and benchmarks for policy-compliant controllability.


{\small
\bibliographystyle{unsrtnat}
\bibliography{main}
}

%%%%%%%%%%%%%%%%%%%%%%%%%%%%%%%%%%%%%%%%%%%%%%%%%%%%%%%%%%%%

\newpage
\appendix

\section{Extended Overview}
\label{appendix:sec:overview}

We include one vocal pop artist (Billie Eilish) and one instrumental piano artist (Ludovico Einaudi) to cover common use cases, and this pairing helps us cover both vocal and instrumental production regimes. The study compares baseline prompts, prompts augmented with artist names, and prompts augmented with name-free descriptors to quantify prompt-level controllability. 

Each artist is represented by fifteen 15-second reference excerpts. Outputs are generated under three conditions: baseline prompts, prompts augmented with the artist name, and prompts augmented with five independent three-token descriptor sets sampled from a large language model; we used GPT-5 in our experiments. Matched seeds are used across all conditions to control for stochastic variability.  

Evaluation is conducted in two embedding spaces: VGGish, following the canonical definition of Fréchet Audio Distance (FAD) \citep{hershey2017vggish,kilgour2019fad}, and CLAP \citep{elizalde2022clap,microsoft2023clap}, a joint audio text model. We report FAD and min-distance attribution in both spaces, and we perform cross-artist validation as a specificity control. Together, these measures capture distributional alignment, per-clip proximity to references, and specificity.

\section{Reproducibility Checklist}
\label{appendix:sec:reproducibility}

Reference excerpts are drawn from publicly available recordings and evaluated only in 15-second segments. 
A manifest of excerpt IDs and timestamps is released, while the audio itself is not redistributed. 
All experiments use the publicly available \texttt{musicgen-small} checkpoint without modification. 
Candidate descriptors are generated automatically by GPT-5 \citep{openai_gpt5_2025}, and the full sampled sets are provided in a manifest file. 

Generation and evaluation scripts are released, with control over prompts and random seeds, as well as embedding extraction. 
Matched seeds are passed directly into the generation script by setting \texttt{torch.manual\_seed(s)} before each run. 
This ensures that, for each seed, one baseline clip, one artist-name clip, and five styled clips (one per descriptor set) are generated under identical random initialization. 
As a result, differences between conditions arise only from the prompt modifiers rather than random noise, enabling paired evaluation.  

Evaluation is conducted in both VGGish and CLAP embedding spaces. 
In VGGish, we report Fréchet Audio Distance (FAD\textsubscript{vgg}) following prior definitions \citep{hershey2017vggish,kilgour2019fad}. 
For CLAP, we use an open-source implementation \citep{elizalde2022clap}, computing analogous FAD\textsubscript{clap}. 
In both spaces we additionally report min-distance attribution as a per-clip metric, using raw median values. 
Results are aggregated across descriptor sets; FAD is reported as mean $\pm$ standard deviation, while min-distance is reported as raw medians.

Released artifacts include descriptors, reference manifests, embeddings, metrics, and figure scripts, enabling exact reproduction of reported results without requiring access to copyrighted recordings.

All experiments were lightweight. Each generation condition (10 seeds × 15 s clips at 32 kHz) ran on a consumer laptop (Apple M2, 24 GB RAM). 
Total compute across all conditions was under 8 CPU hours. 
This modest requirement enables reproduction without specialized hardware or access to GPUs.

\section{Extended Related Work}
\label{appendix:sec:extended-related-work}

\paragraph{Style representations in speech.}  
Global Style Tokens (GST) learn style embeddings in TTS models via a reference encoder, enabling prosody control without explicit supervision \citep{wang2018gst}. We treat GST as a conceptual precedent, while our approach uses training-free, prompt-level interventions for music generation.

\paragraph{Embedding spaces for evaluation.}  
Generative audio evaluation relies heavily on pretrained embeddings. VGGish underpins the canonical definition of Fr'echet Audio Distance (FAD) \citep{hershey2017vggish,kilgour2019fad}, while CLAP provides joint audio text representations used for retrieval and cross-modal tasks \citep{elizalde2022clap}. Recent work shows that metric outcomes depend on embedding choice \citep{tailleur2024fadcorr}, and adaptations of FAD for music evaluation have been proposed \citep{gui2024adaptingfad}. Our study reports results in both VGGish and CLAP to control for embedding choice and provide complementary perspectives.

\section{Detailed Methods}
\label{appendix:sec:detailed-methods}

\subsection{Prompt Construction}  
Candidate descriptors are generated automatically by GPT-5 \citep{openai_gpt5_2025} using a fixed instruction template.  
For each artist, the model is asked to provide a neutral baseline sentence and five independent sets of three short style tokens, subject to constraints on length, format, and coverage of timbre, instrumentation, and mix or rhythm.  
This yields fifteen descriptors per artist, which are concatenated with the baseline to form styled prompts. For clarity, Tables~\ref{tab:prompts-billie} and \ref{tab:prompts-einaudi} list the exact baseline and the five styled prompts per artist.

\begin{table}[t]
\centering
\footnotesize
\setlength{\tabcolsep}{6pt}
\begin{tabularx}{\linewidth}{@{}lX@{}}
\toprule
\textbf{Billie Eilish} & \textbf{Prompt text} \\
\midrule
Baseline &
a moody contemporary pop track with subtle electronic textures, minimal percussion, and an atmospheric groove \\
Styled prompt 1 &
a moody contemporary pop track with subtle electronic textures, minimal percussion, and an atmospheric groove, \underline{breathy lead timbre, sub-bass pulses, dry room reverb} \\
Styled prompt 2 &
a moody contemporary pop track with subtle electronic textures, minimal percussion, and an atmospheric groove, \underline{close-mic lead timbre, sparse percussion, intimate mix space} \\
Styled prompt 3 &
a moody contemporary pop track with subtle electronic textures, minimal percussion, and an atmospheric groove, \underline{airy lead timbre, 808 bass tones, slow sparse groove} \\
Styled prompt 4 &
a moody contemporary pop track with subtle electronic textures, minimal percussion, and an atmospheric groove, \underline{soft lead texture, detuned analog pad, minor-key low tempo} \\
Styled prompt 5 &
a moody contemporary pop track with subtle electronic textures, minimal percussion, and an atmospheric groove, \underline{delicate lead timbre, distorted bass texture, syncopated glitch rhythm} \\
\bottomrule
\end{tabularx}
\caption{\textbf{Billie Eilish prompts.} Baseline and five styled prompts formed by appending the three-token sets.}
\label{tab:prompts-billie}
\end{table}
\begin{table}[t]
\centering
\footnotesize
\setlength{\tabcolsep}{6pt}
\begin{tabularx}{\linewidth}{@{}lX@{}}
\toprule
\textbf{Ludovico Einaudi} & \textbf{Prompt text} \\
\midrule
Baseline &
contemporary instrumental track with gentle dynamics, melodic progression, and subtle harmonic textures \\
Styled prompt 1 &
contemporary instrumental track with gentle dynamics, melodic progression, and subtle harmonic textures, \underline{solo piano texture, repetitive arpeggios, classical reverb} \\
Styled prompt 2 &
contemporary instrumental track with gentle dynamics, melodic progression, and subtle harmonic textures, \underline{delicate piano timbre, string ensemble pad, intimate mix} \\
Styled prompt 3 &
contemporary instrumental track with gentle dynamics, melodic progression, and subtle harmonic textures, \underline{flowing piano lead, simple chord texture, warm acoustic space} \\
Styled prompt 4 &
contemporary instrumental track with gentle dynamics, melodic progression, and subtle harmonic textures, \underline{lyrical piano melody, soft dynamics, contemplative tempo} \\
Styled prompt 5 &
contemporary instrumental track with gentle dynamics, melodic progression, and subtle harmonic textures, \underline{expressive piano tone, minimalist patterns, gentle sustain} \\
\bottomrule
\end{tabularx}
\caption{\textbf{Ludovico Einaudi prompts.} Baseline and five styled prompts formed by appending the three-token sets.}
\label{tab:prompts-einaudi}
\end{table}

In addition, we define an artist-name baseline by appending the artist identifier directly to the baseline description:  
\begin{equation*} 
\texttt{artist\_baseline} = f"\{baseline\} [\{artist\_name\}]"
\end{equation*}  
This provides a control condition where stylization is achieved through the artist label alone. The complete instruction text given to GPT-5 is provided in Appendix~\ref{appendix:sec:prompt}, and the sampled outputs are released to ensure reproducibility.

\subsection{Prompt Template}
\label{appendix:sec:prompt}

The full instruction used to query GPT-5 is shown below. This template was applied identically for each artist, with the artist name inserted in the placeholder.

\begin{lstlisting}[basicstyle=\ttfamily\small,breaklines=true]
You are a veteran music producer and mix engineer. Your job is to produce concise production descriptors that steer the open-source MusicGen model.

TASK
For the artist "{{ARTIST_NAME}}", return a single JSON object containing:
1) "baseline": one neutral, model-friendly sentence for a contemporary pop track suitable for 12-15 s generations (no artist mentions);
2) "sets": exactly 5 diverse "style token" sets, each with 3 short tokens.

CONSTRAINTS
- Tokens are 2-4 words, lowercase ASCII, no proper nouns, no artist names, no lyrics, no brands.
- Avoid tokens that require realistic singing; if needed, describe the lead timbre (e.g., "breathy lead timbre") instead of vocals.
- Each 3-token set must cover: (i) lead timbre/texture, (ii) instrumentation/timbre, (iii) mix/space OR rhythm/harmony.
- The 5 sets must be meaningfully different (no near-duplicates).

OUTPUT (JSON only, no commentary, no markdown):
{
  "artist_name": "{{ARTIST_NAME}}",
  "baseline": "<one-sentence baseline prompt>",
  "sets": [
    ["t1","t2","t3"],
    ["t1","t2","t3"],
    ["t1","t2","t3"],
    ["t1","t2","t3"],
    ["t1","t2","t3"]
  ]
}
\end{lstlisting}

\subsection{Reference Material}  
Reference excerpts are drawn from publicly available recordings, restricted to fifteen-second segments.  
Each excerpt is resampled to 32 kHz to match the generator output.  
The list of excerpt IDs and timestamps is included in the released manifest, but no copyrighted audio is redistributed.  

\subsection{Generation Protocol}  
All experiments use the publicly released \texttt{musicgen-small} checkpoint.  
For each artist, ten random seeds are fixed and rendered with the baseline prompt.  
The same seeds are then used with the artist-name baseline, yielding ten additional outputs.  
Each of the five descriptor sets is also applied to the same seeds, producing fifty styled outputs.  
In total, each artist has seventy outputs (10 baseline, 10 artist-name, 50 styled).  
This matched-seed design controls for randomness across conditions and allows paired comparisons.

\subsection{Evaluation Metrics}  

\textbf{Fr\'echet Audio Distance (FAD).}  
We compute FAD in both VGGish and CLAP embedding spaces, following the canonical definition of \citet{kilgour2019fad}.  
FAD\textsubscript{vgg} provides comparability with prior work, while FAD\textsubscript{clap} incorporates a joint audio text representation.  

\textbf{Min-Distance Attribution.}  
For each generated embedding, we compute its cosine distance to the closest reference embedding.  
This provides a per-clip measure of nearest-neighbor alignment and complements distributional metrics like FAD. We report raw median min-distance values per condition.

\textbf{Embedding Choice.}  
Embedding choice is known to influence evaluation outcomes \citep{tailleur2024fadcorr}. We therefore report results in both VGGish and CLAP \citep{elizalde2022clap,microsoft2023clap}: VGGish supports canonical FAD and comparability with prior work, while CLAP is widely adopted for cross-modal evaluation. Adaptations of FAD for music have also been proposed \citep{gui2024adaptingfad}.

\subsection{Statistical Testing}  
For each artist and condition, metrics are computed over ten matched seeds per descriptor set.  
We aggregate across the five descriptor sets per artist: FAD is reported as mean ± standard deviation; min-distance is reported as raw medians.
All comparisons are paired by seed, ensuring that differences between baseline, artist-name, and styled prompts arise only from the prompt modifiers rather than random variability.  

\subsection{Control Condition}  
For the cross-artist control, all five descriptor sets sampled for one artist are applied to the other artist’s baseline prompt using the same ten seeds. Metrics are computed against the corresponding reference set, and results are averaged across descriptor sets. Degraded alignment in this condition indicates that descriptors encode artist-specific rather than generic stylistic cues.

\section{Ethics, Broader Impacts, and Extended Reproducibility}
\label{appendix:sec:ethics-reproducibility}

\paragraph{Positive societal impacts.}  
This work contributes a reproducible and interpretable framework for studying stylistic controllability in text-to-music systems. By releasing descriptors, manifests, embeddings, and evaluation scripts, we enable other researchers to conduct controlled experiments on prompt-level interventions. The use of human-readable descriptors supports transparency and may inform creative tools for lightweight stylistic steering without retraining or specialized expertise.

\paragraph{Risks and safeguards.}  
As with any generative system, there is potential misuse for unauthorized style imitation. We mitigate this risk by restricting evaluation to short excerpts (15-second), avoiding redistribution of copyrighted audio, and framing alignment quantitatively in embedding space rather than via perceptual mimicry. All released artifacts exclude original recordings and are anonymized where appropriate.

\paragraph{Limitations and ethical considerations.}  
The study is constrained to two artists, short (15-second) clips, and embedding-based metrics, which do not capture long-term musical structure or fully align with perception. Descriptors are sampled from a language model and may reflect biases in phrasing or coverage. Future work should extend to more artists and genres, incorporate listening studies, and test robustness across embedding spaces. This work is an initial step toward policy-compliant controllability in generative music.

\paragraph{Reproducibility.}  
All reference audio is restricted to 15-second segments and is not redistributed. The demo page hosts only generated clips with metadata removed. Released artifacts include code, manifests, embeddings, and generated audio.

\subsection*{Artifacts}
To support reproducibility, we release anonymized artifacts associated with this work:
\begin{itemize}
    \item Code and evaluation scripts: \url{https://github.com/artisticstyles/artisticstyles-neurips}
    \item Website with representative generations: \url{https://artisticstyles.github.io/music-style-control-demo/}
    \item Full generations and reference manifests: \url{https://huggingface.co/datasets/ArtisticStyling/music-style-control-data}
\end{itemize}

\paragraph{License for released assets.} 
All new artifacts introduced in this work (descriptor sets, manifests, embeddings, evaluation outputs, scripts, and demo website) are released under the MIT license. 
This permissive license ensures that other researchers can freely reproduce, adapt, and build upon our results. 

\paragraph{Licenses for external assets.}  
MusicGen-small is released under the CC-BY-NC 4.0 license, VGGish under the Apache-2.0 license,  
Microsoft CLAP under the MIT license, and the reference implementation of Fr\'echet Audio Distance (FAD) under the MIT license.  
GPT-5 was accessed via the ChatGPT web interface in accordance with OpenAI’s Terms of Use.  
Reference audio excerpts are not redistributed; only 15-second segments were used for evaluation under fair-use principles.

\end{document}